# Astro2020 Science White Paper

# Illuminating the dark universe with a very high density galaxy redshift survey over a wide area

**Thematic Areas**:

☑ Cosmology and Fundamental Physics

**Principal Author:**

Name: Yun Wang
Institution: California Institute of technology
Email: wang@ipac.caltech.edu
Phone: (626) 395-1415

**Co-authors:**

Rachel Bean (Cornell), Peter Behroozi (Arizona), Chia-Hsun Chuang (Stanford),
Ian Dell'antonio (Brown), Mark Dickinson (NOAO), Olivier Dore (JPL/Caltech),
Daniel Eisenstein (Harvard), Ryan Foley (UC Santa Cruz), Karl Glazebrook (Swinburne),
Luigi Guzzo (Milan), Christopher Hirata (Ohio State), Shirley Ho (Flatiron),
Michael Hudson (Waterloo), Bhuvnesh Jain (Penn), Priyamvada Natarajan (Yale),
Jeff Newman (Pittsburgh), Alvaro Orsi (CEFCA Spain), Nikhil Padmanabhan (Yale),
John Peacock (Edinburgh), Will Percival (Waterloo), Jason Rhodes (JPL/Caltech),
Eduardo Rozo (Arizona), Lado Samushia (Kansas State), Dan Scolnic (Duke),
Hee-Jong Seo (Ohio), David Spergel (Princeton/Flatiron), Michael Strauss (Princeton),
Risa Wechsler (Stanford), David Weinberg (Ohio State)

**Abstract**:

The nature of **dark energy** remains a profound mystery 20 years after the discovery of cosmic acceleration. A very high number density galaxy redshift survey over a wide area (**HD GRS Wide**) spanning the redshift range of $0.5<z<4$ using the same tracer, carried out using massively parallel wide field multi-object slit spectroscopy from space, will provide *definitive* dark energy measurements with minimal observational systematics by design. The HD GRS Wide will illuminate the nature of dark energy, and lead to revolutionary advances in particle physics and cosmology. It will also trace the cosmic web of dark matter and provide key insight into large-scale structure in the Universe. The required observational facility can be established as part of the *probe* portfolio by NASA within the next decade.

# I. Dark energy: unanswered key question in cosmology

The observational data from recent years have greatly improved our understanding of the Universe. The fundamental questions that remain to be studied in the coming decades include:
***What is the dark energy that is driving the accelerated expansion of the Universe?***

It has been 20 years since cosmic acceleration was discovered (Riess et al. 1998; Perlmutter et al. 1999), yet we are still in the dark about its cause. Dark energy might be an unknown energy component in the Universe, or the consequence of the modification of general relativity (GR). Astro2010 made LSST (optical imaging) and WFIRST (NIR imaging & slitless spectroscopy) top priorities; both will carry out ambitious dark energy projects. DESI (optical spectroscopy) and Euclid (optical/NIR imaging & slitless spectroscopy) will precede and complement these. *These currently planned projects will significantly advance our understanding of the nature of dark energy, but they do not provide definitive measurements for its resolution, due to limits inherent to each.* Given the fundamental importance of discovering the nature of dark energy, Astro2020 should consider future facilities capable of providing definitive data on dark energy.

The precise and accurate measurement of both the cosmic expansion history H(z) and the growth rate of cosmic large-scale structure $f_g(z)$ is required to solve the mystery of dark energy, since modified GR models may predict the same cosmic expansion history as a dark energy model, but would generally predict different growth history of cosmic structure (see e.g., Wang 2008). *A very high number density galaxy redshift survey over a wide area (HD GRS Wide) using space-based slit spectroscopy in the IR is the lowest risk way to obtain definitive measurements on dark energy over the entire relevant redshift range.* The HD GRS Wide minimizes observational systematic errors by design. The theoretical systematic uncertainties (nonlinear structure growth, redshift-space distortions, and the bias between galaxy and matter distributions) can be mitigated through advanced modeling. The HD GRS Wide will provide invaluable constraints on dark matter as well by tracing the cosmic web of dark matter (see Fig.1), and enable the measurement of dark matter filament mass (Epps & Hudson 2017).

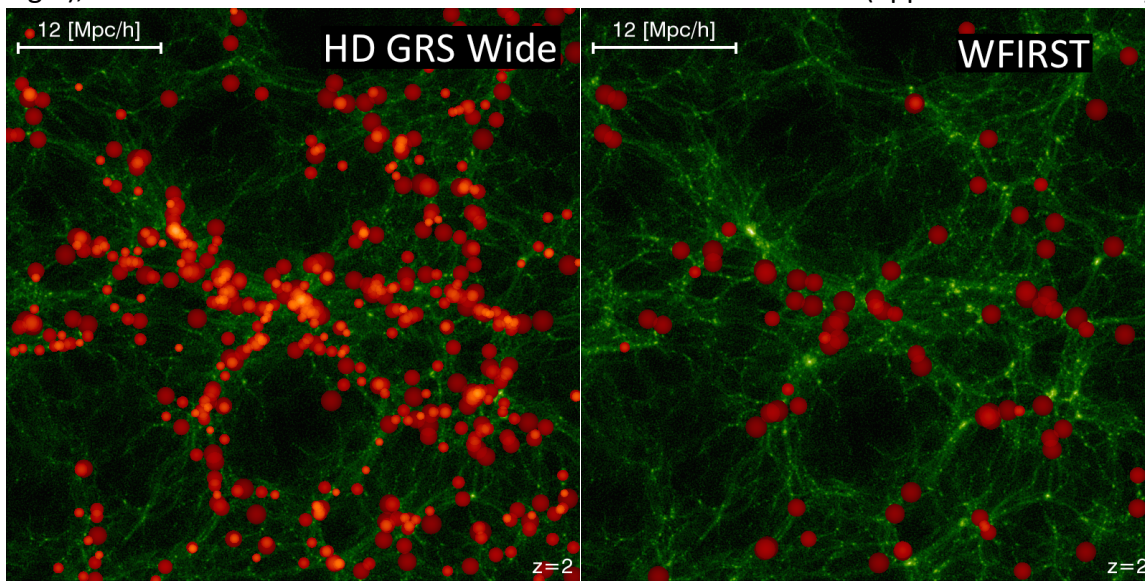

Fig. 1. Cosmic web of dark matter (green) at z=2 traced by galaxies (red filled circles) from an example HD GRS Wide, the ATLAS Wide Survey, which obtains spectra for 70% of galaxies in the WFIRST weak lensing sample, compared to WFIRST GRS. The larger circles represent brighter galaxies. (Wang et al. 2019)

## II. Definitive measurements of dark energy

Given our ignorance of the nature of dark energy, it is critical that we obtain measurements on dark energy that are model-independent (i.e., H(z) & $f_g$(z) as free functions) and definitive (high precision and accuracy) over the entire redshift range over which dark energy influences the expansion of the Universe (i.e., 0<z<4). The measurements at z<1 or even 1.5 can be made from ground-based facilities, thus the focus of the HD GRS Wide should be 1<z<4. The overlap of 0.5<z≲1.5 with ground-based projects is important for cross-check and mitigation of systematic effects. We discuss below the unique advantages of a HD GRS Wide in probing dark energy, using an example survey covering 2000 deg$^2$ at 0.5<z<4, with a galaxy surface number density ~12 times that of the WFIRST GRS and ~50 times that of Euclid, with spectroscopic redshifts for ~183M galaxies. Its most efficient implementation is to obtain spectroscopic redshifts for ~70% of the galaxies from the WFIRST weak lensing sample with Hα emission line flux > 5×10$^{-18}$erg/s/cm$^2$ (see Fig.1 & Fig.2).

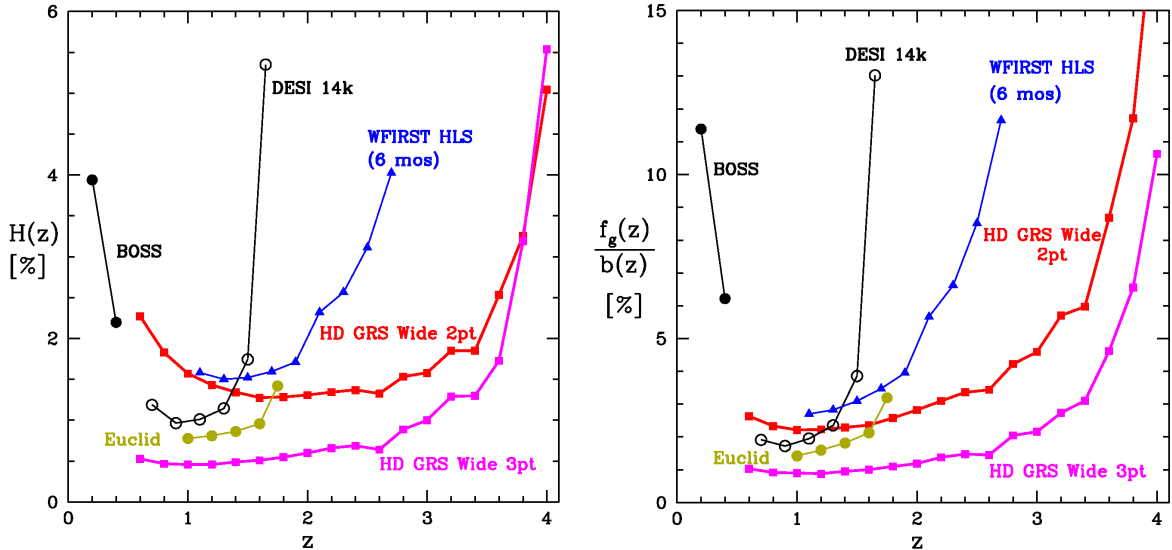

Fig.2. Expected constraints on cosmic expansion history H(z) and growth rate of cosmic large scale structure $f_g$(z) from future surveys. The HD GRS Wide is the same as in Fig.1; "2pt" refers to its power spectrum, "3pt" refers to its bispectrum. Constraints are derived followed the methodology of Wang et al. (2013) & Samushia et al. (2019). For the slitless surveys, we assume an efficiency of 75% for WFIRST (to be launched in 2025), and 40% for Euclid (to be launched in 2022), since the signal-to-noise thresholds for WFIRST and Euclid are 7σ and 3.5σ respectively. The ground-based projects BOSS (completed) and DESI (first light 2019) are complementary to the space-based surveys. The constraints on angular diameter distance $D_A$(z) (not shown in the left panel to avoid crowding the figure) provide a cross-check on H(z). The bias between galaxy and matter distributions is b(z).

**(i) High Number Density and Multi-Tracer BAO/RSD**. Galaxy clustering data from 3D distributions of galaxies is the most robust probe of cosmic acceleration. The baryon acoustic oscillation (BAO) measurements provide a direct measurement of H(z) and angular diameter distance $D_A$(z) (Blake & Glazebrook 2003; Seo & Eisenstein 2003), and the redshift-space distortions (RSD) enable measurement of $f_g$(z) (Guzzo et al. 2008; Wang 2008). The HD GRS Wide provides multiple galaxy tracers of BAO/RSD (red galaxies, different emission-line selected galaxies, and WL shear selected galaxies) over 0.5<z<4, with each at high number densities. These enable robust modeling of BAO/RSD (e.g., the removal of the nonlinear effects via the reconstruction of the linear density field), and significantly tightens constraints on dark energy and modified gravity by evading the cosmic variance when used as multi-tracers (McDonald & Seljak 2009). The HD GRS Wide is needed to study in detail the galaxy-formation systematics (feedback; assembly bias; conformity) that are potential systematics in modeling RSD (Tojeiro et al., 2017). The HD GRS Wide measures H(z), $D_A$(z), and $f_g$(z) over the wide redshift range of 0.5< z < 4 (see Fig.2), with high precision over 0.5<z<3.5, improving the constraints from WFIRST by a factor of three or more at 2<z<3 (beyond the reach of Euclid), and extending constraints to 3<z<4, beyond the reach of both WFIRST and Euclid. If early dark energy remains viable in the 2020s, it can be measured by enhancing the HD GRS Wide with a high z only HD GRS carried out using the same observational facility, targeting Hα emission line galaxies at 3<z<4 selected from WFIRST HLS imaging beyond that of the WL sample.

**(ii) Higher-order Statistics.** The very high number density galaxy samples from the HD GRS Wide provide the ideal data set for studying higher-order statistics of galaxy clustering. For a galaxy sample with number density n, the shot noise scales as 1/n for galaxy power spectrum (2pt), and $1/n^2$ for galaxy bispectrum (3pt). Fig.2 shows that *the HD GRS Wide 3pt galaxy clustering gives definitive measurements on dark energy, outperforming all other measurements* (Samushia et al. 2019)**.** Since the 3pt statistics provides information not contained in the 2pt statistics, the combination of these is needed to optimally extract the cosmological information from galaxy clustering data (see, e.g., Gagrani & Samushia 2017), and enables the direct measurement of bias b(z). In addition, the cross-3pt function, galaxy-galaxy-lensing shear, will help break degeneracies between galaxy bias and cosmological parameters. It will be measured with high signal-to-noise for the sample sizes discussed here. While the use of galaxy clustering 2pt statistics is now standard in cosmology, the use of the 3pt statistics is still limited due to a number of technical challenges (see, e.g., Yankelevich & Porciani 2019). *A HD GRS Wide in the next decade will take advantage of the anticipated future advances in galaxy 3pt statistics.*

**(iii) 3D WL with Spectroscopic Redshifts.** The HD GRS Wide replaces photometric redshifts with spectroscopic redshifts for ~70% of the lensed galaxies in the WFIRST HLS WL sample. This eliminates the photo-z calibration ladder as a major source of systematic uncertainty in WL. Furthermore, one can identify every pair of galaxies that are near each other in 3D space for 70% of the sample, dramatically suppressing the systematic error from intrinsic galaxy alignments. This improves the measurement of the growth rate index (used to parametrize the gravity model) by ~50% (Wang et al. 2019).

**(iv) Joint Analysis of Weak Lensing and Galaxy Clustering.** Most galaxies in the WFIRST WL sample will not have WFIRST spectroscopy; but ~70% will have spectroscopy at z > 0.5 from the HD GRS Wide. This super data set of both WL shear and 3D galaxy clustering data for the same 183M galaxies over 0.5<z<4 enables a straightforward joint analysis of the data. This facilitates the precise modeling of the bias between galaxies and matter, b(z), and provides the ultimate measurement of H(z) and $f_g$(z). We expect this to result in definitive measurements of dark energy and modified gravity for 0.5<z<4, with the removal of photo-z errors as the primary systematic for weak lensing, and detailed modeling of bias systematic for BAO/RSD.

**(v) Type Ia Supernovae (SNe Ia).** The HD GRS Wide can easily include the host galaxy redshifts of nearly all 30,000 SNe Ia that will have lightcurves measured by the LSST Deep Drilling Fields and WFIRST SN surveys, over the redshift range of 0.2<z<2.0. This will provide a powerful measurement of H(z) that is independent of cosmic large scale structure. The WFIRST and LSST

SN surveys will cover ~40 deg$^2$ each. Current plans for follow-up spectroscopy will acquire redshifts for ≲ 10% of SN host galaxies at z>1, leading to a small and largely biased sample. Multiple studies (e.g. Sullivan et al. 2010) have shown that there is a relation between SN luminosity and host galaxy properties, so a large effort should be made to create an unbiased sample of host galaxies to as high a redshift as possible. With host galaxy redshifts up to z=2.0 from the HD GRS Wide, the statistical distance precision of WFIRST and LSST SNe should be sub-1% out to z=2.0 (Hounsell et al. 2018). This is comparable to the constraints shown in Fig.2 from the HD GRS Wide measurement errors from BAO/RSD. Without the host galaxy redshifts, SN analyses will be forced to rely on photometric redshifts, which for the full SN sample degrades the cosmological precision by introducing a series of systematic uncertainties and reducing the statistical precision as well.

**(vi) Clusters.** The abundance of mass-calibrated galaxy clusters and cluster strong lensing cosmography provide complementary measurements of cosmic expansion history and growth rate of large-scale structure (Jullo et al. 2010). The HD GRS Wide will include spectroscopy for the 40,000 clusters with M>$10^{14} M_\odot$ expected to be found by WFIRST HLS imaging (Spergel et al. 2015). This cluster sample will have mass accurately measured by the deep 3D WL data, and cluster membership precisely determined by spectroscopic redshifts. The HD GRS Wide also provides independent cluster mass function evolution measurements using infall amplitude rather than lensing, which would eliminate degeneracies that affect weak lensing and clusters the same way in LSST data. This and stacked cluster strong lenses will provide robust and powerful constraints on dark energy models and their potential evolution, offering an important crosscheck to constraints from BAO/RSD and 3D WL, and a unique data set of spectroscopic clusters for studying cluster astrophysics.

**(vii) Voids.** Cosmic voids are underdense patches of the Universe. Voids can be used to test for modifications of gravity from GR: in many models modifications are stronger where the matter density is low. They can also be used to make geometrical measurements using the fact that the line-of-sight and transverse scales of voids should be equivalent on average - i.e. there is no preferred direction in the Universe. The HD GRS Wide provides an ideal data set for probing cosmology with voids: the dense sampling means that we will pick up many galaxies close to the void centers, which contain the most information. In addition, contrasting biased tracers across environments of different densities, spanning voids to clusters, could provide valuable tests of gravity (Valogiannis & Bean 2018, 2019). For a detailed discussion of voids as a cosmological probe, see Astro2020 science white paper by Pisani et al. (2019).

### III. Weighing dark-matter-dominated filaments in the cosmic web

A key prediction of the cold dark matter model is the existence of the "cosmic web", a network of low-density filaments connecting dark matter halos. *Spectroscopic redshifts* are required to locate the galaxy groups and clusters that are connected by filaments in redshift space. The uncertainty associated with photometric redshifts scatters the true physical pairs and smears the filamentary structure. Epps & Hudson (2017) demonstrated that the stacked dark-matter-dominated filament connecting pairs of massive galaxies can be detected when spectroscopic redshifts along with weak lensing data are available for most of the source galaxies lensed by the massive galaxy pairs. Clampitt et al. (2016) made similar findings. The HD GRS Wide and WFIRST imaging together provide the ideal data set for such detections (see

Fig.3). The detection of these filaments enabled by the HD GRS Wide over a significant cosmic volume will test the cold dark matter model for structure formation, and provide key insight into large-scale structure in the Universe. Finally, the filament lensing can be compared with filament signals in SZ surveys to probe dark matter, a subject of ongoing work.

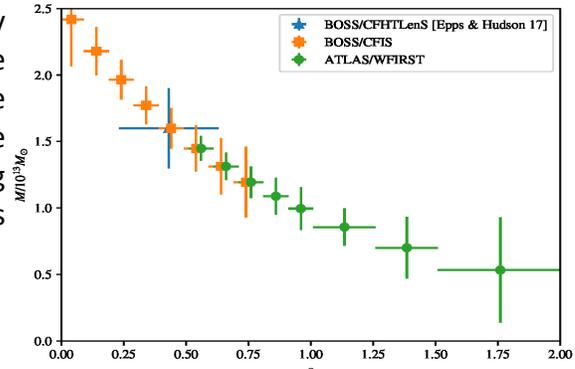

Fig.3. Conservative estimate of the measurement errors of filament mass from a HD GRS Wide, ATLAS Wide, for a 6-10 Mpc/h cosmic web filament between $10^{13}$ M$_\odot$ halos. (Wang et al. 2019)

## IV. Observational facility required

The HD GRS Wide for obtaining definitive measurements on dark energy would require massively parallel wide field slit spectroscopy in the IR from space, to remove the loss of sensitivity due to the punishing telluric effects on the ground and the observational systematics intrinsic to space-based slitless spectroscopy (a primary concern for Euclid/WFIRST). It should achieve very high number density of galaxies using multi-object slit spectroscopy with multiplex factor >5000, spectral resolution $R = \lambda/\Delta\lambda \gtrsim 600$ to resolve all primary spectral lines, and cover a wavelength range of 1-4 μm to ensure the observation of multiple emission lines.

ATLAS Probe is an example observational facility that will be able to execute the HD GRS Wide in 1.6 years of observing time, as the Wide tier of 3 nested surveys for studying galaxy evolution (Wang et al. 2019). ATLAS Probe is the spectroscopic follow-up mission to WFIRST, with a 1.5m aperture telescope and a field of view of 0.4 deg$^2$. It uses Digital Micro-mirror Devices as slit selectors for R=1000 multi-object slit spectroscopy, with a multiplex factor ∼ 6000. ATLAS Probe is designed to fit within the NASA probe-class space mission cost envelope.

Performing a high-density redshift survey from the ground would permit similar types of analyses at $0<z\lesssim1.5$. By covering 15-20K deg$^2$, such surveys can achieve similar volume to 2000 deg$^2$ at z>1, giving more evolutionary baseline and maximizing our leverage at low redshift. Such efforts are starting with the DESI Bright Galaxy Sample (z<0.5) and PFS surveys, providing a strong platform for methodological development to exploit the full information of high density sampling. Future ground surveys (perhaps with new facilities) could extend this mapping to a scope of $10^8$ galaxies.

The HD GRS Wide will produce a spectacular and unprecedented dark energy and legacy science data set of ∼183M galaxy spectra. To maximize its science return, we expect that data centers and archives will require advanced facilities such as server-side analysis.

## V. Recommendation

Given the fundamental importance of solving the mystery of dark energy, it is important for the community to invest resources in obtaining definitive measurements on dark energy with minimal observational systematics by design. A very high number density wide area galaxy redshift survey (HD GRS Wide) spanning the redshift range of 0.5<z<4 using the same tracer, carried out using massively parallel wide field multi-object slit spectroscopy from space, will provide such measurements (see Figs.1-2) that can illuminate the nature of dark energy, and lead to revolutionary advances in particle physics and cosmology. The required observational facility can be established as part of the probe portfolio by NASA within the next decade.